\documentclass{ifacconf}

\usepackage{graphicx}      
\usepackage{natbib}        

\usepackage{amsmath,amssymb,amsfonts}
\usepackage{algorithmic}
\usepackage{graphics}
\usepackage{textcomp}
\usepackage{xcolor}

\usepackage{graphicx}  
\usepackage{subcaption} 

\graphicspath{{figure/}}

\newtheorem{definition}{Definition}[section]

\begin{document}
\begin{frontmatter}

\title{Autonomous Cooperative Levels of Multiple-Heterogeneous Unmanned Vehicle Systems\thanksref{footnoteinfo}} 

\thanks[footnoteinfo]{This research was supported by Unmanned Vehicles Core Technology Research and Development Program through the National Research Foundation of Korea (NRF) and Unmanned Vehicle Advanced Research Center (UVARC) funded by the Ministry of Science and ICT, the Republic of Korea (No.: NRF-2020M3C1C1A02086425).}
\thanks[footnoteinfo]{Note that this paper is included in the web proceedings of Joint Conference of APCATS, AJSAE, AAME 2023.}

\author[First]{Yoo-Bin Bae} 
\author[Second]{Yeong-Ung Kim} 
\author[Second]{Jun-Oh Park}
\author[Second]{Hyo-Sung Ahn}

\address[First]{Aeronautics Research Directorate Unmanned Aircraft System Research Division, Korea Aerospace Research Institute (KARI), 
   Daejeon 34133, Republic of Korea (e-mail: ybbae@kari.re.kr).}
\address[Second]{School of Mechanical Engineering, Gwangju Institute of Science and Technology (GIST), 
  Gwangju 61005, Republic of Korea \newline(e-mail: \{yeongungkim, junoingist\}@gm.gist.ac.kr, hyosung@gist.ac.kr)}

\begin{abstract}                
As multiple and heterogenous unmanned vehicle systems continue to play an increasingly important role in addressing complex missions in the real world, the need for effective cooperation among unmanned vehicles becomes paramount. The concept of autonomous cooperation, wherein unmanned vehicles cooperate without human intervention or human control, offers promising avenues for enhancing the efficiency and adaptability of intelligence of multiple-heterogeneous unmanned vehicle systems. Despite the growing interests in this domain, as far as the authors are concerned, there exists a notable lack of comprehensive literature on defining explicit concept and classifying levels of autonomous cooperation of multiple-heterogeneous unmanned vehicle systems. In this aspect, this article aims to define the explicit concept of autonomous cooperation of multiple-heterogeneous unmanned vehicle systems. Furthermore, we provide a novel criterion to assess the technical maturity of the developed unmanned vehicle systems by classifying the autonomous cooperative levels of multiple-heterogeneous unmanned vehicle systems.
\end{abstract}


\end{frontmatter}

\section{Introduction}
A networked system composed of multi-agents has recently been extensively studied due to its applicability to various fields (Ref. \cite{oh2015survey, olfati2007consensus}). The aim of these studies is to perform high-level missions by networking multiple low-level and inexpensive agents, and this technology of networked system can be widely applied, for example, military operations using multi-drones, traffic control, and energy distribution. In particular, multiple and heterogeneous unmanned vehicle systems are playing an increasingly important role as a solution to complex missions since different types of vehicles in heterogeneous systems can perform missions well suited to their characteristics.
With the development of wireless communication, on-board data processing, and on-board sensing technologies, the unmanned vehicle systems are rapidly becoming intelligent. An important concept related to the system intelligence is autonomy. In general, autonomy of a single unmanned vehicle means that the vehicle can observe, analyze, decide, and act on its own without human intervention or human control, and many studies related to this concept have been conducted by several research organizations, such as Defense Advanced Research Projects Agency (DARPA), Air Force Research Laboratory (AFRL), National Institute of Standards and Technology (NIST), and National Aeronautics and Space Administration (NASA) (Ref. \cite{chung2017offensive, huang2005framework}).

\begin{figure}[h!]
\begin{center}
\includegraphics[width=1\columnwidth]{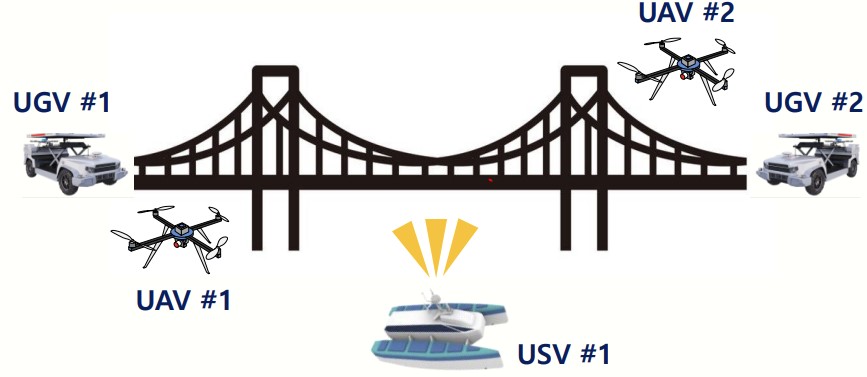}
\caption{An Example of Multiple-Heterogeneous Unmanned Vehicle Systems (UAV-UGV-USV).}
\label{line}
\end{center}
\end{figure}

Unlike a single unmanned vehicle, multiple unmanned vehicle systems require effective cooperation between vehicles to perform complex missions, and the degree of the cooperation is gradually increasing. In this aspect, the concept of autonomy of cooperation, i.e., autonomous cooperation, is required in case of multiple-heterogeneous unmanned vehicle systems. Despite the growing interests on these networked systems, as far as the authors are concerned, there is a lack of comprehensive literature on defining an explicit concept of autonomous cooperation and classifying autonomous cooperative levels of the networked systems. Note that the classifying the autonomous cooperative levels of these systems is paramount to assess the technical maturity of the developed unmanned vehicle systems. Also, this technical classification provides a clear picture of the functional requirements of networked systems suitable for each level of autonomous cooperation. As a consequence, to this end, this article aims to define the explicit concept of autonomous cooperation. Furthermore, a novel criterion is provided for the assessment of the developed networked systems by classifying the autonomous cooperative levels of multiple-heterogeneous unmanned vehicle systems.
This article is organized as follows. In Section 2, we first define several terminologies related to automation, autonomy, and cooperation of the systems. After that, we introduce some relating works on autonomy levels of unmanned vehicle systems. In Section 3, we present the concept of autonomous cooperation. Furthermore, autonomous cooperation levels are classified, which is the main result of this article. Lastly, all results are summarized in Conclusions.

\section{Preliminaries}

\subsection{Related Term Definition}\label{sec2.1}
Before we study the concept of autonomous cooperation of multiple-heterogeneous unmanned vehicle systems, we clearly define several terminologies related to automation and autonomy of systems as follows.

\begin{enumerate}

\item Automation and Autonomy of a System
\newline
a) Automatic system: A system that performs pre-determined procedures and makes no decision without human intervention.
\newline
b) Autonomous system: A system that makes decisions according to the specific algorithms without human intervention considering the external environment and real-time status.
\newline
c) Intelligent system: A system that learns principles and makes decisions without human intervention by considering the external environment and real-time status.

\item Automation and Autonomy of a Single Unmanned Vehicle System
\newline
a) Automatic unmanned vehicle system: A system that moves by following pre-determined way points and trajectories.
\newline
b) Autonomous unmanned vehicle system: A system that makes decisions and moves without human intervention considering the external environment and real-time status.

\item Automation and Autonomy of Cooperation of a Multiple Unmanned Vehicle System
\newline
a) Automatic cooperative system: A system that a group of unmanned vehicles cooperate together in a pre-determined way to perform a mission. 
\newline
b) Autonomous cooperative system: A system that a group of unmanned vehicles cooperate by themselves to perform a mission without human intervention by considering the external environment and sharing real-time status between vehicles.

\end{enumerate}

\subsection{Related Works}\label{sec2.2}
This subsection introduces some existing classification method.

\begin{enumerate}

\item General Autonomy Level
\begin{table}[h!]
\captionof{table}{}
\centering
\setlength{\tabcolsep}{1pt}
\resizebox{\columnwidth}{!}{
\begin{tabular}{|p{2cm}|p{2cm}|p{2cm}|}
\hline
Level & Definition & Description \\
\hline
1 & Human in the loop & Human control \\
\hline
2 & Human on the loop & Autonomous action under human monitoring  \\
\hline
3 & Human out of the loop & Autonomous action without human \\
\hline
\end{tabular}}
\label{table1}
\end{table}
\newline
Based on Table \ref{table1}, many methods have been studied according to the degree of human intervention/control or human-computer interaction (Ref. \cite{sheridan1992telerobotics, national2005autonomous, suresh2010role}). These classification methods are general but not specific, therefore, it is difficult to apply these methods to the systems with various technical elements.

\item  Autonomy Levels for Unmanned Systems Framework, ALFUS (Ref. \cite{huang2005framework}).
\begin{figure}[h!]
\begin{center}
\includegraphics[width=0.9\columnwidth]{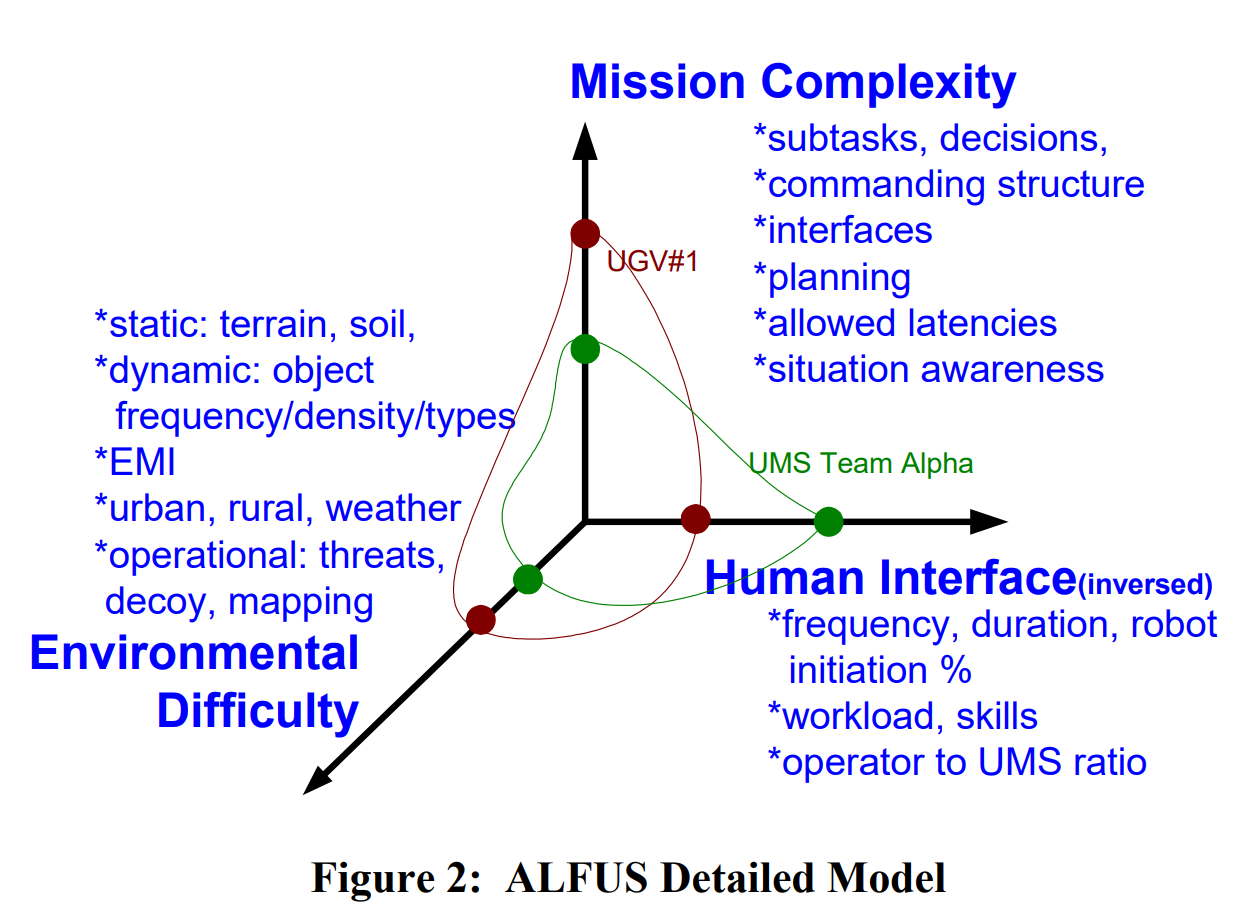}
\caption{The 3-Axis model of ALFUS. Image is extracted from \cite{huang2005framework}.}
\label{figure2}
\end{center}
\end{figure}

\begin{figure}[h!]
\begin{center}
\includegraphics[width=0.9\columnwidth]{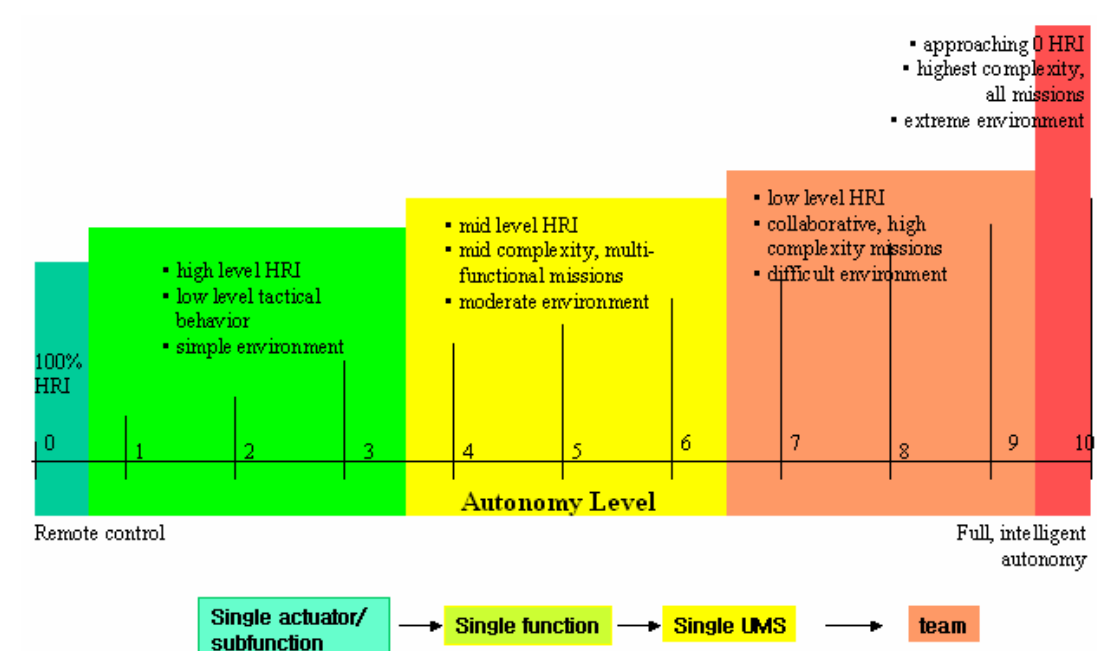}
\caption{Autonomy Level Classification in ALFUS. Image is extracted from \cite{huang2005framework}.}
\label{figure3}
\end{center}
\end{figure}
Different from the general classification methods of autonomy levels as stated in Section 2.2.(1)., the authors in \cite{huang2005framework} identified three autonomy keywords: 1) mission complexity, 2) environment difficulty, 3) human interface. Based on these keywords, they presented autonomy levels for unmanned systems framework, ALFUS, as shown in Fig. \ref{figure2} and Fig. \ref{figure3}. This method is detailed and systematic compared to the other classification methods. 

\end{enumerate}

\subsection{Autonomy of Cooperation}
As stated in Section \ref{sec2.2}, several autonomy level classification methods have been studied for a single unmanned vehicle system. However, as far as the authors are concerned, there is no comprehensive literature on defining concept and classifying levels of autonomous cooperation of multiple-heterogeneous unmanned vehicle systems. In this aspect, the following section studies on the concept of autonomous cooperation of multiple-heterogeneous unmanned vehicle systems.

\section{What is Autonomous Cooperation?}\label{sec3}

\subsection{Concept of Autonomous Cooperation}
From the term definition in Section \ref{sec2.1}, we recall the definition of the autonomous cooperation of networked systems as follows.
\begin{definition}[Autonomous Cooperation]\label{Def1}
Autonomous \newline cooperation of multiple-heterogeneous unmanned vehicle systems means that a group of unmanned vehicles cooperate by themselves to perform a mission without human intervention by considering the external environments and sharing real-time status between vehicles.
\end{definition}

Definition \ref{Def1} implies full autonomy of cooperation of networked systems, and it is necessary to classify the autonomy levels of cooperation to reach full autonomy.

\subsection{Autonomous Cooperative Levels}
Autonomy keywords are required to classify autonomous cooperative levels of multiple-heterogeneous unmanned vehicle systems. To identify the autonomy keywords, let us consider the following cooperation process of humans: 

There is a group of people to perform a teamwork mission by working together. To solve a teamwork mission, 

(A) They categorize a mission into smaller tasks (Mission-Task categorization).

(B) They check each limitation and human-task compatibility (Ability recognition).

(C)	All tasks are allocated to people (Task allocation).

(D)	They determine the order of allocated tasks (Task arrangement).

(E)	If there are additional tasks created or left, they reallocate tasks (Task flexibility).

Inspired from the above cooperative dynamics observed in human interactions, we have identified five pivotal keywords for autonomy cooperation. Based on these keywords, the following table shows the new autonomous cooperative levels of multiple-heterogeneous unmanned vehicle systems, which is the main result of this article. 

\begin{table}[h!]
\captionof{table}{}
\resizebox{\columnwidth}{!}{
\begin{tabular}{|p{1.5cm}|p{1.5cm}|p{1.5cm}|p{1.5cm}|}
\hline
 {} & Human in the loop (1) & Human on the loop (2) & Human out of the loop (3) \\
\hline
Mission-Task categorization (A) & A1 & A2 & A3 \\
\hline
Ability recognition (B) & B1 & B2 & B3  \\
\hline
Task allocation (C) & C1 & C2 & C3 \\
\hline
Task arrangement (D) & D1 & D2 & D3 \\
\hline
Task flexibility (E) & E1 & E2 & E3 \\
\hline
\end{tabular}}
\label{table2}
\end{table}
The following cooperation scenario is provided to help better understanding of the presented autonomous cooperative levels in the above table.

\textbf{A Scenario:}

\begin{figure}[h!]
\begin{center}
\includegraphics[width=1\columnwidth]{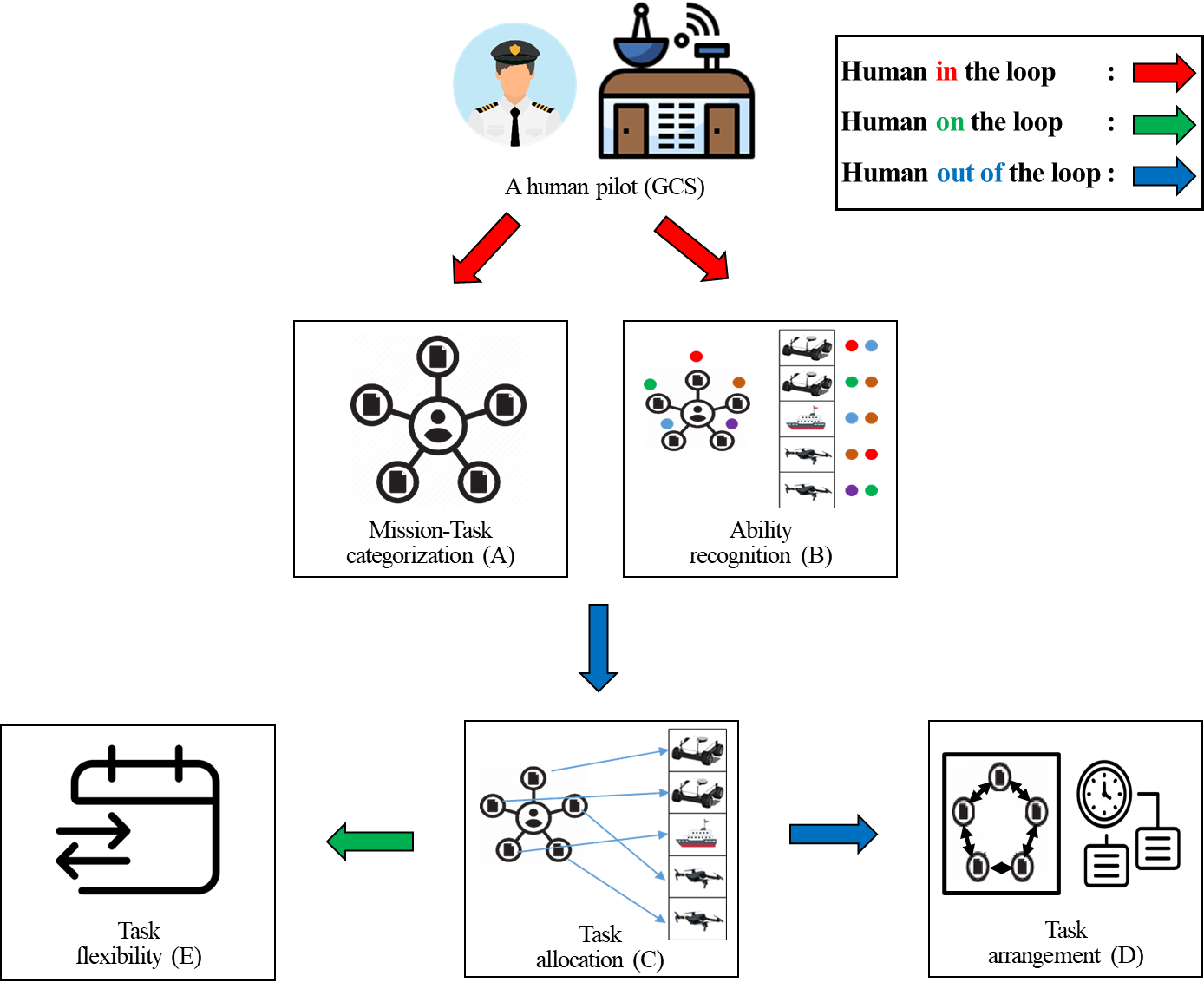}
\caption{Autonomous cooperation of the system in a scenario.}
\label{figure4}
\end{center}
\end{figure}

In Fig. \ref{figure4}, a human pilot breaks down a mission into several tasks (A1) and checks the task-vehicle compatibility requirements (B1) in a ground control station (GCS). After each vehicle receives information about categorized tasks, task limitation, and task-vehicle compatibility from the GCS, all vehicles decide tasks through auction algorithms (C3) and arrange the order of tasks (D3) in a fully distributed way. If additional tasks are created, the tasks are allocated to the specific vehicles under the monitoring of the human pilot (E2). 
Utilizing the autonomous cooperative levels as delineated in Table \ref{table2}, the autonomous cooperative level attributed to a given scenario's system is determined as A1-B1-C3-D3-E2. This elucidation consequently underscores the imperative of augmenting the technical proficiency in the realms of mission-task categorization (A), ability recognition (B), and task flexibility (E) in order to elevate the autonomous cooperative level of the system.
Also, we note that the proposed method in Table 2 allows for the distinction of a comprehensive range of 243(3*3*3*3*3) autonomous cooperative levels.


\section{Conclusion}
This article delves into an exploration of the concept of autonomous cooperation within systems. Additionally, we undertake the task of classifying the distinct levels of autonomous cooperation exhibited by multiple-heterogeneous unmanned vehicle systems. The authors anticipate that this endeavor will serve as a fundamental cornerstone, laying the groundwork for forthcoming investigations into autonomous cooperation within diverse networked systems. As part of our future research endeavors, we intend to delve deeper into the classification of autonomous cooperative levels inherent to various networked systems. Moreover, a critical aspect of our forthcoming work involves the identification and delineation of the technological requisites corresponding to each autonomous cooperative level.
\begin{ack}
This research was supported by Unmanned Vehicles Core Technology Research and Development Program through the National Research Foundation of Korea (NRF) and Unmanned Vehicle Advanced Research Center (UVARC) funded by the Ministry of Science and ICT, the Republic of Korea (No.: NRF-2020M3C1C1A02086425).

 Also, we note that this paper is included in the web proceedings of Joint Conference of APCATS, AJSAE, AAME 2023
\end{ack}

\bibliography{mybib}             
                                                   







\end{document}